\documentclass{aa}  

\usepackage{graphicx}
\usepackage{longtable}
\usepackage{txfonts}
\begin{document}

   \title{Searching for $\lambda$ Boo stars in binary systems}

   
   \titlerunning{$\lambda$ Boo stars in binary systems}
   \authorrunning{Alacoria et al.}

   \author{J. Alacoria\inst{1,5}, C. Saffe\inst{1,2,5}, A. Collado\inst{1,2,5}, A. Alejo\inst{1,2,5},
           D. Calvo\inst{1,2,5}, P. Miquelarena\inst{1,2,5}, E. Gonz\'alez\inst{2},
           M. Flores\inst{1,2,5}, M. Jaque Arancibia\inst{3,4},
            \and F. Gunella\inst{1,5}
           }

\institute{Instituto de Ciencias Astron\'omicas, de la Tierra y del Espacio (ICATE-CONICET), C.C 467, 5400, San Juan, Argentina.
         \and Universidad Nacional de San Juan (UNSJ), Facultad de Ciencias Exactas, F\'isicas y Naturales (FCEFN), San Juan, Argentina.
         \and Instituto de Investigaci\'on Multidisciplinar en Ciencia y Tecnolog\'ia, Universidad de La Serena, Ra\'ul Bitr\'an 1305, La Serena, Chile
         \and Departamento de F\'isica y Astronom\'ia, Universidad de La Serena, Av. Cisternas 1200 N, La Serena, Chile.         
        \and Consejo Nacional de Investigaciones Cient\'ificas y T\'ecnicas (CONICET), Argentina
         }

   \date{Received xxx, xxx ; accepted xxxx, xxxx}

 
  \abstract
   {The origin of $\lambda$ Boo stars is currently unknown.
    Very few of them are known as members of multiple systems, which could provide benchmark laboratories to study their origin.}
   {Our goal is to find new candidate $\lambda$ Boo stars that belong to binary systems.
    Then, a detailed abundance determination of some candidates could confirm their true $\lambda$ Boo nature, 
    while the composition of eventual late-type companions could be used as a proxy for the initial composition of the $\lambda$ Boo stars. 
   }
   {We cross-matched a homogeneous list of candidate $\lambda$ Boo stars with a recent Gaia eDR3 catalog of resolved binaries. 
    Then, we carried out a detailed abundance determination via spectral synthesis of three of these systems,
    hosting a candidate $\lambda$ Boo star and a late-type companion:
    HD 98069 + UCAC4 431-054639, HD 87304 + CD-33 6615B and HD 153747 + TYC 7869-2003-1.
    Stellar parameters were estimated by fitting observed spectral energy distributions (SEDs)
    with a grid of model atmospheres using the online tool VOSA, together with Gaia eDR3 parallaxes and the
    PARAM 1.3 interface.
   Then, the abundances were determined iteratively for 31 different species by fitting synthetic spectra using the SYNTHE program
   together with local thermodynamic equilibrium (LTE) ATLAS12 model atmospheres.
   Specific opacities were calculated for each star, depending on its arbitrary composition and microturbulence velocity,
   v$_\mathrm{micro}$, through the opacity sampling (OS) method. 
   The abundances of the light elements C and O were corrected by non-LTE effects.
   The complete chemical patterns of the stars were then compared to those of $\lambda$ Boo stars.}
   {
   We obtained a group of 19 newly identified binary systems which contain a candidate $\lambda$ Boo star,
   allowing to $\sim$duplicate the number of $\lambda$ Boo stars currently known in multiple systems.
   This important group could be used in further studies of $\lambda$ Boo stars.
   For the first time, we performed a detailed abundance analysis of three of these binary systems
   which includes a candidate $\lambda$ Boo star and a late-type companion.
   We confirmed the true $\lambda$ Boo nature of the three early-type stars (HD 87304, HD 98069 and HD 153747),
   and obtained mostly a solar-like composition for their late-type components.
   Adopting as a proxy the late-type stars,
   we showed that the three $\lambda$ Boo stars were initially born with a solar-like composition. 
   This is an important constraint for any scenario trying to explain the origin of $\lambda$ Boo stars.
   The present work provides three numerical examples of possible "starting" and "ending" compositions
   to test formation models of $\lambda$ Boo stars.
   Also, the solar-like composition of the late-type stars supports the idea
   that $\lambda$ Boo stars are Population I objects, however, we caution that other explanations are also possible.
   }
   {We performed, for the first time, a detailed abundance analysis of binary systems including 
   a $\lambda$ Boo star and a late-type companion. We obtained a solid indication that $\lambda$ Boo stars
   born from a solar-like composition and established an important constraint to test formation models
   of $\lambda$ Boo stars.
   }
   

   \maketitle
%

\section{Introduction}

The $\lambda$ Boo stars are considered as one of the long-standing puzzles in astrophysics.
They are a rare class of stars, making up about $\sim$2\% of the population of stars of spectral type A \citep{gray-corbally98,paunzen01b}.
The main characteristic of $\lambda$ Boo stars is a notable surface depletion of most Fe-peak elements
together with near-solar abundances of lighter elements C, N, O, and S \citep[e.g. ][]{kamp01,andrievsky02,heiter02,alacoria22}.
The origin of the $\lambda$ Boo peculiarity remains a challenge despite recent efforts
\citep[see e.g. ][]{jura15,kama15,murphy-paunzen17,kunitomo18,saffe21,alacoria22}.
The scenarios trying to explain the notable abundance pattern point to diverse ideas, including
acretion of gas from a circumstellar disk \citep{venn-lambert90},
the ingestion of comets and volatile-rich objects \citep{gray-corbally02},
the interaction of the star with the interstellar medium (ISM) or with a diffuse interstellar cloud \citep{cowley82,kamp-paunzen02,mg09},
and the ablation of volatile gases from a near hot-Jupiter planet \citep{jura15}.
However, none of the suggested scenarios seem to explain the origin of the $\lambda$ Boo stars
\citep[see e.g. the recent discussion in ][]{murphy-paunzen17,alacoria22}.

The detection of $\lambda$ Boo stars as members of binary sytems is considered an important finding,
being a laboratory to test physical conditions under which these stars formed.
However, currently few $\lambda$ Boo stars are clearly identified as members of these systems.
For example, \citet{paunzen12a,paunzen12b} identified a $\sim$dozen of $\lambda$ Boo stars as members of early-type binary systems,
and proposed a method to test the accretion scenario. They suggested that two early-type stars passing through a diffuse cloud should display,
in principle, the same superficial peculiarity \citep[see also ][]{alacoria22}.
The detection of a binary or multiple system including a $\lambda$ Boo star and a late-type companion is also valuable.
In this case, the chemical composition of the late-type component could be considered as a proxy of the original composition
from which both stars formed, being crucial to test formation models of $\lambda$ Boo stars \citep[e.g. ][]{alacoria22}.
If the composition of the $\lambda$ Boo and the late-type star differs significantly,
this would be a solid indication that the $\lambda$ Boo was originally born with a different composition.
Also, the "differential pattern" between these stars could help to precisely quantify
the $\lambda$ Boo phenomena, that is, to measure element by element the effect produced by the $\lambda$ Boo peculiarity.
However, to our knowledge, only one multiple system including a late-type component is reported in the literature: 
the remarkable triple system HD 15165 \citep{alacoria22}.
The examples mentioned show that the detection of multiple systems including a $\lambda$ Boo component
could provide benchmark laboratories, being an important tool to study the origin of $\lambda$ Boo stars.

Progress in understanding the $\lambda$ Boo stars has been hindered
by a somewhat heterogeneous literature \citep[see e.g. section 1.2 in ][]{murphy15}.
This motivated to reevaluate the membership of 
previously reported $\lambda$ Boo stars \citep[see ][]{murphy15,gray17,murphy20}, 
using mainly classification spectroscopy.
Together, these three works comprise a complete and homogeneous sample of predominantly southern $\lambda$ Boo stars.
On the other hand, \citet{el-badry21} recently obtained an extensive catalog of 1.3 (1.1) millon of spatially resolved binaries
with a bound probability $>$90\% ($>$99\%) within $\sim$1 kpc of the Sun, using Gaia eDR3 data.
We caution that 15\% of A-type stars have companions with periods of 100-1500 days \citep{murphy18}, 
which would be undetected in the \citet{el-badry21} study.
Then, with the aim of detect $\lambda$ Boo stars in multiple systems,
we cross-matched the compilation of 118 $\lambda$ Boo stars with this recent catalog of resolved binaries.
The results of this experiment are presented in the Table \ref{table.LamBoo.binaries},
showing a list of $\lambda$ Boo stars together with their corresponding binary companions.
The columns present the binary number, star name, V magnitude, coordinates J2016.0 ($\alpha$ and $\delta$),
proper motions ($\mu_{\alpha}$ and $\mu_{\delta}$), parallax $\pi$, and finally separation (in seconds and au).
As previously explained, this important group of newly identified binary systems could be used
in further studies of $\lambda$ Boo stars.
Most $\lambda$ Boo stars of Table \ref{table.LamBoo.binaries} have late-type companions.
The binaries include separations ranging between $\sim$150 au up to $\sim$57000 au.
We note that the list includes one binary system (the number 7, HD 198160/HD 198161) having a 
previously known $\lambda$ Boo star candidate \citep{gray88,sturenburg93,alacoria22}.


We present in this work, 
the first detailed analysis of binary systems which include a candidate $\lambda$ Boo star and a late-type companion.
We focused on three relatively bright binary systems included in the Table \ref{table.LamBoo.binaries},
obtaining fundamental parameters and abundances for their components.
The separation of the stars allowed us to obtain clean individual spectra,
free from a possible contamination from its stellar companion.
This will allow us to determine, for the first time, a solid proxy for the starting chemical composition
of the $\lambda$ Boo stars, that is, the chemical composition from which the $\lambda$ Boo stars born.
This is a critical constraint for any model that explains the origin of $\lambda$ Boo stars.
In addition, the stars analyzed together with those presented in the Table \ref{table.LamBoo.binaries},
are important laboratories for further studies of $\lambda$ Boo stars.
The new binary systems reported in Table \ref{table.LamBoo.binaries} allow to $\sim$duplicate 
the number of $\lambda$ Boo stars currently known in multiple systems.
The sample could also help to determine if the presence of an stellar companion could play a role
in the development of the peculiarity.

This work is organized as follows. In Sect. 2, we describe the observations and data reduction.
In Sect. 3, we present the stellar parameters and chemical abundance analysis. 
In Sect. 4, we show the results and discussion. 
Finally, in Sect. 5, we highlight our main conclusions.

\begin{table*}
\centering
\caption{Visual binary systems that include a $\lambda$ Boo star component (the first star listed), identified with Gaia eDR3 data.}
\scriptsize
\begin{tabular}{clrrrrrrrr}
\hline
Binary & Star      & V     & $\alpha$    & $\delta$     & $\mu_{\alpha}$ & $\mu_{\delta}$ & $\pi$ & Sep. & Sep. \\
number & name      &       & J2016.0     & J2016.0      & [mas/yr]       & [mas/yr]       & [mas] & ["]  & [au] \\
\hline
1 & HD 112948 & 9.35 & 13 0 42.78 & -36 9 5.83 & -27.528 & -2.159 & 2.719 & 24.38 & 8967.79 \\ 
   & Gaia DR3 6154835886339634304 &    & 13 0 44.58 & -36 9 16.56 & -27.447 & -2.160 & 2.769 &  &  \\ 
2 & HD 91130 & 5.91 & 10 31 51.40 & +32 22 46.53 & 16.652 & 7.876 & 12.101 & 38.87 & 3211.74 \\ 
   & Gaia DR3 748463991961311488 &   & 10 31 48.36 & +32 22 41.11 & 16.302 & 8.891 & 12.187 &  &  \\ 
3 & HD 107233 & 7.36 & 12 19 55.73 & -48 18 59.76 & 50.923 & -17.777 & 12.379 & 15.02 & 1213.63 \\ 
   & Gaia DR3 6130179235015300096 &   & 12 19 54.22 & -48 18 58.99 & 50.085 & -18.487 & 12.332 &  &  \\ 
4 & HD 319 & 5.92 & 0 7 46.98 & -22 30 31.37 & 68.586 & -35.437 & 11.690 & 1.84 & 157.43 \\ 
   & HD 319B &    & 0 7 47.06 & -22 30 32.87 & 69.557 & -26.724 & 11.399 &  &  \\ 
5 & HD 153747 & 7.42 & 17 2 53.84 & -38 27 36.60 & 7.488 & 13.836 & 5.647 & 322.64 & 57131.24 \\ 
   & TYC 7869-2003-1 & 11.84 & 17 2 38.75 & -38 23 7.05 & 7.604 & 13.963 & 5.627 &  &  \\ 
6 & HD 168740 & 6.12  & 18 25 31.63 & -63 1 19.17 & -0.017 & -102.405 & 14.056 & 65.11 & 4631.9 \\ 
   & Gaia DR3 6630272406475726592 &   & 18 25 40.04 & -63 1 50.19 & -0.193 & -104.575 & 14.086 &  &  \\ 
7 & HD 198160 & 6.21 & 20 51 38.70 & -62 25 46.34 & 82.885 & -47.171 & 13.599 & 2.44 & 179.57 \\ 
   & HD 198161 &   & 20 51 39.05 & -62 25 45.93 & 82.063 & -42.345 & 13.625 &  &  \\ 
8 & HD 46722 & 9.29 & 6 33 46.53 & -27 57 21.84 & -18.649 & 0.201 & 4.590 & 40.55 & 8833.78 \\ 
   & Gaia DR3 2895732444923489408 &   & 6 33 49.46 & -27 57 10.07 & -18.687 & 0.199 & 4.573 &  &  \\ 
9 & HD 168947 & 8.11 & 18 24 30.07 & -44 11 56.67 & 8.242 & 0.875 & 3.872 & 104.64 & 27022.47 \\ 
   & Gaia DR3 6721717074879155840 &   & 18 24 22.47 & -44 10 51.32 & 8.286 & 0.511 & 3.815 &  &  \\ 
10 & HD 40588 & 6.19 & 6 1 10.02 & +31 2 4.27 & -20.388 & 0.307 & 13.342 & 29.91 & 2241.98 \\ 
   & Gaia DR3 3449874691731245184 &   & 6 1 8.07 & +31 1 47.90 & -21.721 & -0.135 & 13.468 &  &  \\ 
11 & HD 75654 & 6.36 & 8 49 52.27 & -39 8 29.12 & -63.354 & 40.784 & 13.914 & 284.08 & 20417.17 \\ 
   & Gaia DR3 5525846343983205376 &   & 8 50 7.45 & -39 4 46.68 & -62.674 & 40.545 & 13.941 &  &  \\ 
12 & HD 87304 & 9.49 & 10 3 23.43 & -33 41 2.58 & -19.776 & 0.833 & 2.204 & 6.97 & 3160.67 \\ 
   & CD-33 6615B & 12.60 & 10 3 23.05 & -33 40 57.41 & -19.808 & 0.824 & 2.165 &  &  \\ 
13 & HD 98069 & 8.16 & 11 16 44.62 & -3 56 34.57 & -0.435 & -4.224 & 3.420 & 55.24 & 16148.71 \\ 
   & UCAC4 431-054639 & 11.18 & 11 16 48.07 & -3 56 54.40 & -0.331 & -4.303 & 3.286 &  &  \\ 
14 & HD 94390 & 8.96 & 10 53 7.81 & -44 49 13.53 & -15.059 & -1.990 & 3.737 & 7.98 & 2136.26 \\ 
   & Gaia DR3 5387858145792881152 &   & 10 53 8.56 & -44 49 14.24 & -14.327 & -2.293 & 3.713 &  &  \\ 
15 & HD 162193 & 8.66 & 17 53 33.95 & -59 42 25.73 & 1.441 & -15.944 & 3.549 & 15.28 & 4305.24 \\ 
   & Gaia DR3 5917808942063785088 &   & 17 53 34.53 & -59 42 40.36 & 1.896 & -15.852 & 3.569 &  &  \\ 
16 & HD 223352 & 4.57 & 23 48 55.67 & -28 7 50.68 & 99.435 & -106.240 & 22.730 & 74.71 & 3287.04 \\ 
   & HD 223340 & 9.28 & 23 48 50.61 & -28 7 17.35 & 96.589 & -105.274 & 22.682 &  &  \\ 
17 & HD 94326 & 7.76 & 10 52 33.30 & -46 13 1.95 & -24.595 & 8.099 & 3.426 & 40.31 & 11766.79 \\ 
   & Gaia DR3 5363470118895488512 &   & 10 52 32.35 & -46 13 41.04 & -24.233 & 8.412 & 3.563 &  &  \\ 
18 & BD-15 4515 & 9.97 & 17 19 12.91 & -15 54 37.47 & 4.679 & 0.519 & 3.064 & 62.43 & 20374.60 \\ 
   & Gaia DR3 4136170951949470464 &   & 17 19 10.80 & -15 55 31.96 & 4.725 & 0.211 & 3.028 &  &  \\ 
19 & HD 34799 & 8.23 & 5 18 46.06 & -28 52 20.98 & 3.349 & -15.314 & 4.291 & 20.72 & 4829.66 \\ 
   & Gaia DR3 2954518486936209792 &   & 5 18 45.12 & -28 52 4.34 & 3.262 & -15.948 & 4.150 &  &  \\ 
\hline
\end{tabular}
\normalsize
\label{table.LamBoo.binaries}
\end{table*}

\section{Observations}

We present in Table \ref{table.parallax} the visual magnitude V, coordinates, proper motions,
parallax and signal-to-noise per pixel (@5000 \AA), for the stars studied in this work.
The spectral data of the binary system HD 87304 + CD-33 6615B were acquired through the
Gemini High-resolution Optical SpecTrograph (GHOST),
which is attached to the 8.1 m Gemini South telescope at Cerro Pach\'on, Chile.
GHOST is illuminated via 1.2" integral field units that provide the
input light apertures. The spectral coverage of GHOST between
360-900 nm is appropriate for deriving stellar parameters and
chemical abundances using several features. 
It provides a high resolving power R$\sim$50000 in the standard resolution 
mode\footnote{https://www.gemini.edu/instrumentation/ghost}.
The read mode was set to medium, as recommended for relatively bright targets.
The observations were taken on October 10, 2024 and October 24, 2024 (PI: Carlos Saffe, Program ID: GS-2024B-Q-403)
using the same spectrograph configuration for both stars.
The exposure times were 120 sec and 1100 sec (for HD 87304 and CD-33 6615B),
obtaining a final signal-to-noise ratio (S/N) between $\sim$220-275 per pixel measured at $\sim$5000 {\AA} for both stars.
The spectra were reduced using the GHOST data reduction pipeline v1.1.0, which works under 
DRAGONS\footnote{https://www.gemini.edu/observing/phase-iii/reducing-data/dragons-data-reduction-software}. This is
a platform for the reduction and processing of astronomical data.

\begin{table*}
\centering
\caption{Magnitudes and astrometric data for the stars studied in this work.}
\begin{tabular}{lcccccccc}
\hline
Star        & V     & $\alpha$    & $\delta$     & $\mu_{\alpha}$ & $\mu_{\delta}$ & $\pi$   & Spectra & S/N       \\
            &       & J2000       & J2000        & [mas/yr]       & [mas/yr]       & [mas]   &         & @5000 \AA \\
\hline
HD 87304         & 9.49  & 10 03 23.43  & -33 41 02.58 & -19.776 & 0.833  & 2.204 & Gemini+GHOST & 275 \\ 
CD-33 6615B      & 12.60 & 10 03 23.05  & -33 40 57.41 & -19.808 & 0.824  & 2.165 & Gemini+GHOST & 220 \\ 
HD 98069         & 8.16  & 11 16 44.62  & -03 56 34.57 & -0.435  & -4.224 & 3.420 & CASLEO+REOSC & 350 \\ 
UCAC4 431-054639 & 11.18 & 11 16 48.07  & -03 56 54.40 & -0.331  & -4.303 & 3.286 & CASLEO+REOSC & 170 \\ 
HD 153747        & 7.42  & 17 02 53.84  & -38 27 36.60 & 7.488   & 13.836 & 5.647 & CASLEO+REOSC & 300 \\ 
TYC 7869-2003-1  & 11.84 & 17 02 38.75  & -38 23 07.05 & 7.604   & 13.963 & 5.627 & CASLEO+REOSC & 130 \\ 
\hline
\end{tabular}
\normalsize
\label{table.parallax}
\end{table*}

The spectra of the binary systems HD 98069 + UCAC4 431-054639 and HD 153747 + TYC 7869-2003-1
were obtained at the Complejo Astr\'onomico El Leoncito (CASLEO) during
Mar 15-16, 2024, Jun 11-12, 2024 and Jul 11-12, 2024 (PI: Jos\'e Alacoria, Program ID: JS-2024A-06). 
We used the \emph{Jorge Sahade} 2.15-m telescope equipped with a REOSC echelle
spectrograph\footnote{On loan from the Institute d'Astrophysique de Liege, Belgium.} and a SOPHIA 
2048$\times$2048 (152-VS-X eXcelon) CCD detector.
The REOSC spectrograph uses gratings as cross-dispersers. We used a grating with 300 lines mm$^{-1}$ (dispersor \# 270),
which provides a resolving power of $\sim$ 13500 or more, covering appropriately a spectral range of $\lambda\lambda$3700--7500.
Five individual spectra for each object were obtained and then combined, 
reaching an average S/N per pixel of $\sim$235 measured at $\sim$5000 \AA.
We take the stellar spectra for each target followed by a ThAr lamp in order to derive
an appropriate pixel versus wavelength solution.
The data were reduced with Image Reduction and Analysis Facility (IRAF) 
following the standard recipe for echelle spectra (i.e., bias and flat corrections, 
order-by-order normalization, scattered light correction, etc.).
The
 continuum normalization and other operations (such as Doppler
correction and combining spectra) were performed using IRAF.

\section{Stellar parameters and abundance analysis}

Stellar parameters were determined as homogeneously as possible
for the stars in our sample, including both late-type and early-type stars.
We used the Virtual Observatory Sed Analyzer\footnote{http://svo2.cab.inta-csic.es/theory/vosa/} \citep[VOSA, ][]{bayo08}
and the spectral energy distributions (SEDs) constructed from photometric data, 
to obtain the stellar effective temperatures (T$_\mathrm{eff}$) of the objects in our sample
via atmospheric model fitting.
Observed SEDs were unreddened by VOSA using the extinction maps of 
\citet{schlegel98} and following the procedure of \citet{bilir08} to derive A$_{v}$.
We used a grid of Kurucz-NEWODF models by \citet{kurucz93} covering T$_\mathrm{eff}$ between 3500 K and 13000 K
with a step of 250 K.
Then, we performed a Bayesian estimation of surface gravities $\log g$ using Gaia eDR3 parallaxes with
the PARAM 1.3 interface\footnote{http://stev.oapd.inaf.it/cgi-bin/param$\_$1.3} \citep{dasilva06}.
Temperatures and gravities derived for the stars in our sample are presented in the Table \ref{table.params}.

\begin{table*}
\centering
\caption{Fundamental parameters derived for the stars in this work.}
\begin{tabular}{lccccr}
\hline
Star             & T$_{\rm eff}$  & $\log g$        & [Fe/H]           & v$_\mathrm{micro}$ & $v\sin i$     \\
                 &  [K]           &  [dex]          & [dex]            & [km s$^{-1}$]      & [km s$^{-1}$] \\
\hline
HD 87304         & 7750 $\pm$ 250 & 3.85 $\pm$ 0.06 & -1.17 $\pm$ 0.14 & 3.21 $\pm$ 0.80    & 140.5 $\pm$ 3.6 \\
CD-33 6615B      & 6250 $\pm$ 250 & 4.34 $\pm$ 0.06 & -0.05 $\pm$ 0.15 & 1.02 $\pm$ 0.25    &  13.7 $\pm$ 0.7 \\
HD 98069         & 7500 $\pm$ 250 & 3.67 $\pm$ 0.06 & -1.05 $\pm$ 0.13 & 3.00 $\pm$ 0.75    &  80.5 $\pm$ 2.1 \\
UCAC4 431-054639 & 6000 $\pm$ 250 & 4.17 $\pm$ 0.09 & -0.16 $\pm$ 0.17 & 0.68 $\pm$ 0.17    &  21.0 $\pm$ 0.8 \\
HD 153747        & 9250 $\pm$ 250 & 4.14 $\pm$ 0.04 & -0.94 $\pm$ 0.16 & 2.37 $\pm$ 0.59    &  82.1 $\pm$ 2.7 \\
TYC 7869-2003-1  & 5750 $\pm$ 250 & 4.50 $\pm$ 0.03 & +0.00 $\pm$ 0.15 & 0.42 $\pm$ 0.11    &   10.3 $\pm$ 0.2 \\
\hline
\end{tabular}
\normalsize
\label{table.params}
\end{table*}

Projected rotational velocities $v\sin i$ were first estimated by fitting the line \ion{Mg}{II} 4481.23 \AA\
and then refined by fitting most \ion{Fe}{I} and \ion{Fe}{II} lines in the spectra. 
Synthetic spectra were calculated using the program SYNTHE \citep{kurucz-avrett81} together with ATLAS12 \citep{kurucz93} model atmospheres.
and then convolved with a rotational profile (using the Kurucz's command \textit{rotate}) and 
with an instrumental profile for each spectrograph (using the command \textit{broaden}).
The resulting $v\sin i$ values are shown in the 6th column of Table \ref{table.params},
covering between 10.3 $\pm$ 0.2 km s$^{-1}$ and 140.5 $\pm$ 3.6 km s$^{-1}$ for the stars in our sample.
Microturbulence velocity v$_\mathrm{micro}$ was estimated as a function of T$_{\rm eff}$ following the
formula of \citet{gebran14}, which is valid for $\sim$6000 K $<$ T$_{\rm eff}$ $<$ $\sim$10000 K.
We adopted an uncertainty of $\sim$25 $\%$  for v$_\mathrm{micro}$, as suggested by \citet{gebran14},
and then this uncertainty was taken into account in the abundance error calculation.

We applied an iterative procedure to determine the chemical abundances for the stars in our sample.
As a first step, we computed an ATLAS12 \citep{kurucz93} model atmosphere, adopting initially solar
abundances from \citet{asplund09}. The corresponding abundances were then obtained by fitting
the observed spectra with the program SYNTHE \citep{kurucz-avrett81}.
With the new abundance values, we derived a new model atmosphere and restarted the process again.
In each step, opacities were calculated for an arbitrary composition and v$_\mathrm{micro}$ using the opacity
sampling (OS) method, similar to previous works \citep{saffe20,saffe21,saffe22,alacoria22}.
In this way, parameters and abundances were consistently derived using specific opacities rather than solar-scaled values,
until reach the same input and output abundances \citep[for more details, see ][]{saffe21}.
Possible differences originating from the use of solar-scaled opacities instead of an arbitrary
composition were recently estimated for solar-type stars \citep{saffe18,saffe19}.
These differences could become particularly important when modeling chemically peculiar stars,
where solar-scaled models could result in a very different atmospheric structure \citep[e.g., ][]{piskunov-kupka01}
and reach abundance differences up to 0.25 dex \citep{khan-shulyak07}.

\begin{figure}
\centering
\includegraphics[width=8cm]{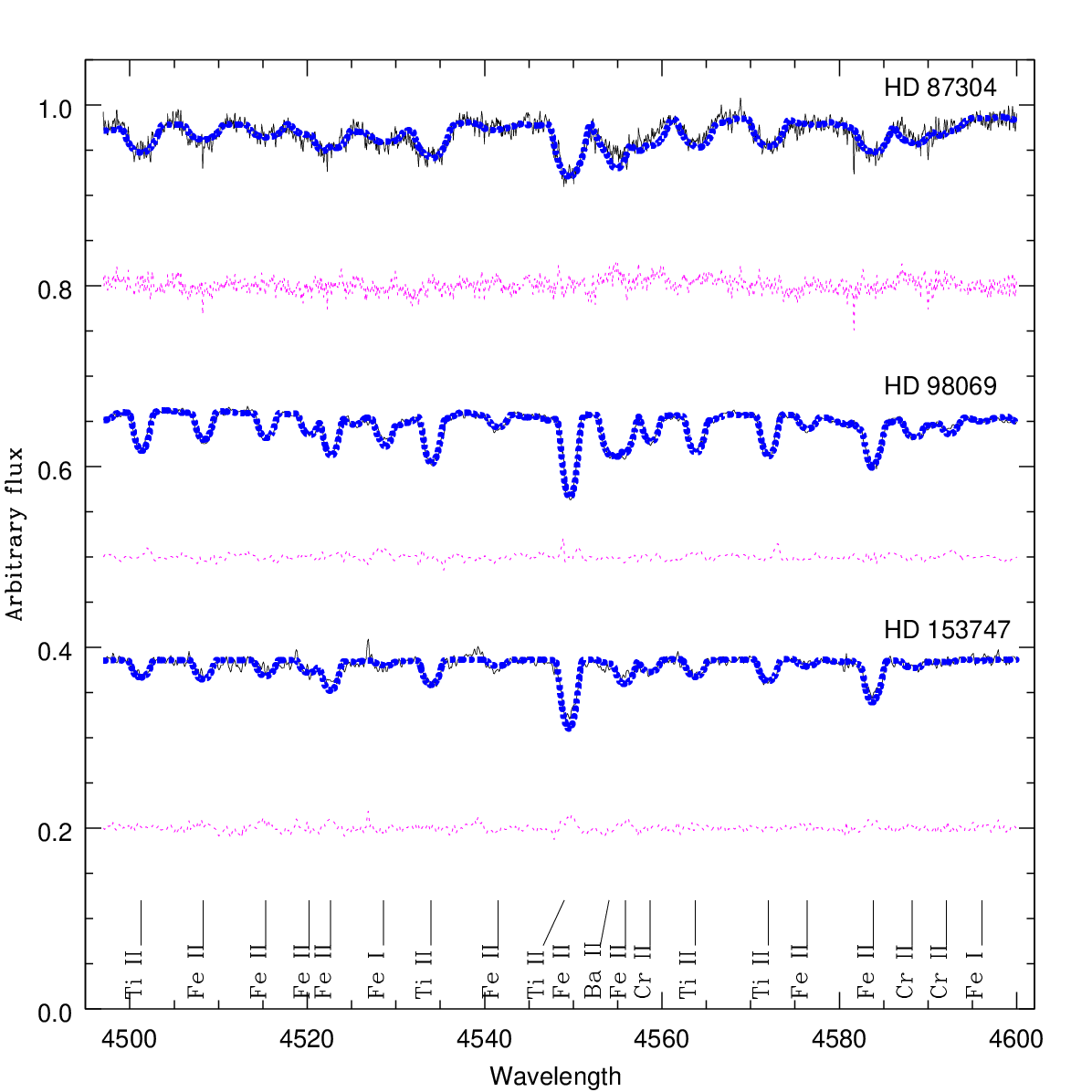}
\caption{Observed, synthetic, and difference spectra (black, blue dotted, and magenta lines) 
for some stars in our sample.}
\label{fig.region2}%
\end{figure}

We derived the chemical abundances for 31 different species, including 
\ion{Li}{I}, \ion{C}{I}, \ion{O}{I}, \ion{Na}{I}, \ion{Mg}{I}, \ion{Mg}{II},
\ion{Al}{I}, \ion{Si}{I}, \ion{Si}{II}, \ion{S}{II}, \ion{Ca}{I}, \ion{Ca}{II}, \ion{Sc}{I}, \ion{Sc}{II},
\ion{Ti}{I}, \ion{Ti}{II}, \ion{V}{I}, \ion{Cr}{I}, \ion{Cr}{II}, \ion{Mn}{I},
\ion{Fe}{I}, \ion{Fe}{II}, \ion{Co}{I}, \ion{Ni}{I}, \ion{Cu}{I},
\ion{Zn}{I}, \ion{Sr}{II}, \ion{Y}{II}, \ion{Ba}{II}, \ion{La}{II} and \ion{Ce}{II}.
The atomic line list and laboratory data used in this work are the same described in \citet{saffe21}.
Figure \ref{fig.region2} presents an example of observed, synthetic, and difference spectra 
(black, blue dotted, and magenta lines) for some stars in our sample.
There is a good agreement between the results of modeling and the observations for the lines of different chemical species.

The uncertainty in the abundance values was estimated considering different sources.
We estimated the measurement error, e$_{1}$, from the line-to-line dispersion
as $\sigma/\sqrt{n}$, where $\sigma$ is the standard deviation and n is the number of lines.
For elements with only one line, we adopted for $\sigma$ the standard deviation of the iron lines.
Then, we determined the contribution to the abundance error due to the uncertainty 
in stellar parameters. We modified T$_{\rm eff}$, $\log g$, and v$_\mathrm{micro}$ by their
uncertainties and recalculated the abundances, obtaining the corresponding
differences e$_{2}$, e$_{3}$, and e$_{4}$ (we adopt a minimum of 0.01 dex for these errors).
Finally, the total error e$_{tot}$ was estimated as the quadratic sum of e$_{1}$, e$_{2}$, e$_{3}$, and e$_{4}$.
The abundances with their total error e$_{tot}$, the individual errors e$_{1}$ to e$_{4}$, and the
number of lines n, are presented in the Tables \ref{tab.abunds.HD87304} to \ref{tab.abunds.TYC7869} of the Appendix.

\subsection{NLTE effects}

In the case of $\lambda$ Boo stars, light-element non-local thermodynamic equilibrium (NLTE)
abundances are particularly important.
For example, an average \ion{O}{I} correction of -0.5 dex was derived by \citet{paunzen99}
for a sample of $\lambda$ Boo stars; while for \ion{C}{I}, they estimated an average correction of -0.1 dex.
\citet{rentzsch96} derived neutral carbon NLTE abundance corrections by using a multilevel 
model atom for stars with T$_\mathrm{eff}$ between 7000 K and 12000 K, log g between
3.5 and 4.5 dex, and metallicities from -0.5 dex to +1.0 dex.
She showed that \ion{C}{I} NLTE effects are negative (calculated as NLTE-LTE) and
depend basically on equivalent width W$_{eq}$.
Near $\sim$7000 K, the three lower levels of \ion{C}{I} are always in LTE; however, increasing
the T$_\mathrm{eff}$ values increases the underpopulation of these levels respect to LTE
by UV photoionization.
Thus, we estimated NLTE abundance corrections of \ion{C}{I} for the early-type stars in our sample
by interpolating in their Figs. 7 and 8 as a function of T$_\mathrm{eff}$, W$_{eq}$,
and metallicity. We applied a similar correction in previous works \citep{alacoria22},
which allows the comparison of abundance values.

NLTE abundance corrections for \ion{O}{I} were derived by \citet{sitnova13},
who used a multilevel model atom with 51 levels.
The authors showed that NLTE effects lead to an strengthening of \ion{O}{I} lines,
producing a negative NLTE correction.
They calculated NLTE corrections for a grid of model atmospheres,
including stars with spectral
types from A to K (T$_\mathrm{eff}$ between 10000 and 5000 K).
We estimated NLTE abundance corrections of \ion{O}{I} (IR triplet 7771 \AA)
for the stars in this work, interpolating based on Table 11 of \citet{sitnova13}, as
a function of T$_\mathrm{eff}$. We note that other \ion{O}{I} lines present corrections 
lower than $\sim$-0.02 dex \citep[see, e.g., Table 5 of ][]{sitnova13}.

\subsection{Comparisons with the literature}

\begin{figure}
\centering
\includegraphics[width=8.0cm]{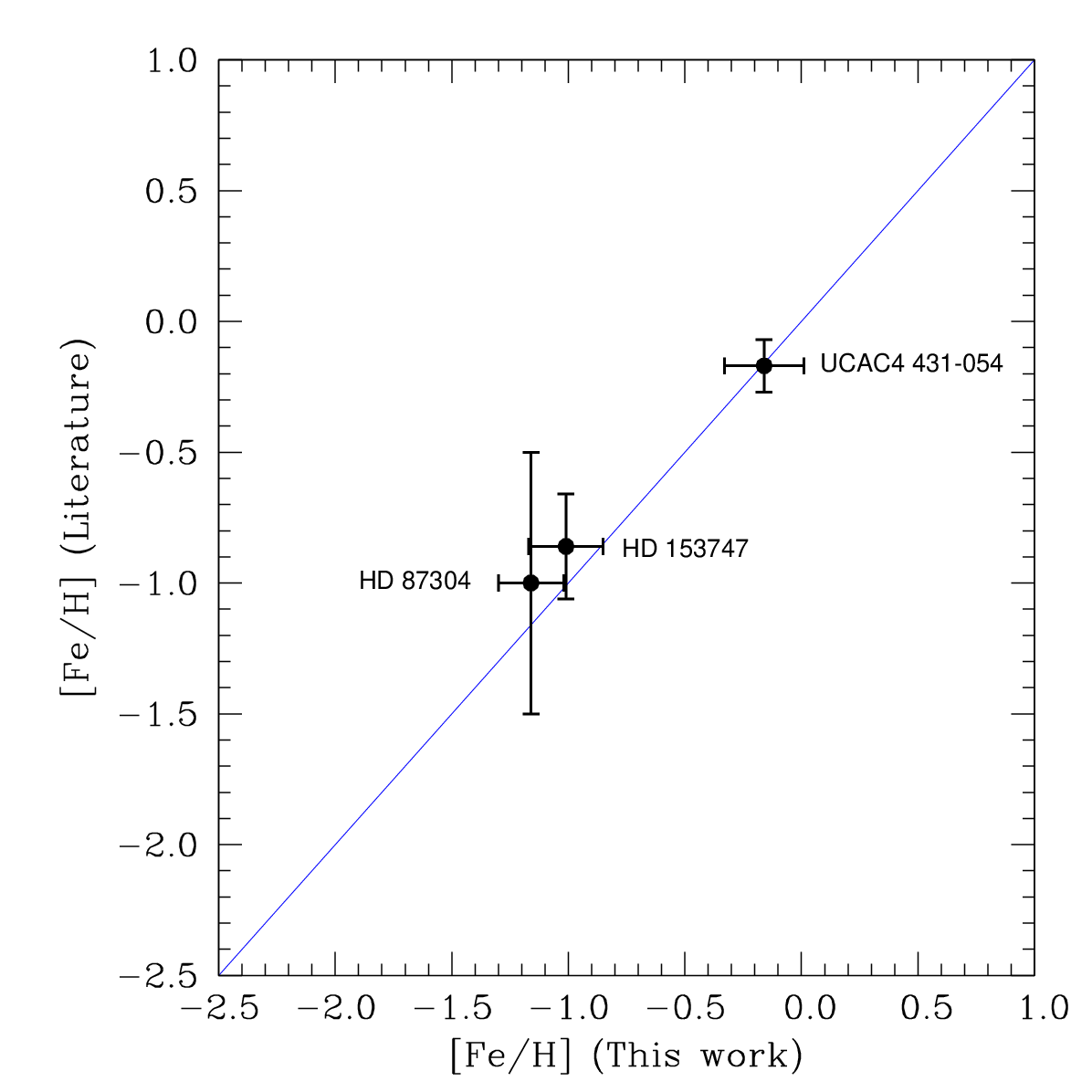}
\caption{Comparison of [Fe/H] values derived in this work with those from literature.}
\label{fig.metal.liter}
\end{figure}

We present in Fig. \ref{fig.metal.liter} a comparison of [Fe/H] values
derived in this work, with those taken from literature for the stars
UCAC4 431-054639 \citep{steinmetz20},
HD 87304 \citep{gray17} and HD 153747 \citep{paunzen02b}.
In general, there is a good agreement with the literature.
The two stars with the lowest metallicity (HD 87304 and HD 153747) seem to present 
literature values slightly higher than in the present work.
However, the values could still be considered similar within their error bars.

\section{Discussion}

We discuss in this section the chemical composition of the binary systems,
which include a candidate $\lambda$ Boo star and a late-type companion.
The chemical patterns of the early-type stars are compared to an average pattern
of $\lambda$ Boo stars. 
We caution that to derive an average pattern for $\lambda$ Boo stars is not an easy task,
due to the relatively low number of stars homogenously analyzed
\citep[see e.g. Section 4.1 of ][]{alacoria22}.
We use in this work the same average chemical pattern of \citet{alacoria22}.
Basically, we adopt the data derived by \citet{heiter02}, who homogeneously determined
abundances for a number of $\lambda$ Boo stars,
and then we excluded from the average those stars without CNO values, similar to \citet{alacoria22}.

\subsection{Binary system HD 87304 + CD-33 6615B}



The star HD 87304 was classified by \citet{gray17} as A8 V kA2.5mA2.5 $\lambda$ Boo.
However, the authors caution that the identification as member of the $\lambda$ Boo class
should be followed up with high-resolution abundance studies to confirm a $\lambda$ Boo abundance pattern,
similar to the present work.
The candidate $\lambda$ Boo star HD 87304 is accompanied by CD-33 6615B,
a late-type star separated by 6.97 arcsec or 3160.67 au \citep{el-badry21}.
\citet{el-badry21} estimated empirically a probability that each pair is a chance alignment 
(R in their paper and "R\_chance\_align" in their catalog).
The authors consider "high bound probability" or "high confidence" pairs those with R$<$0.1,
corresponding to $>$ 90\% probability of being bound.
In particular, this pair presents R$=$1.24 10$^{-4}$, being considered as a high bound probability pair.
The separation allows us to analyze both stars independently without contamination from its companion,
which is different than other $\lambda$ Boo stars \citep[see e.g.][]{paunzen12a,paunzen12b}.
To our knowledge, there is no detailed abundance determination for any component of this binary system.

\begin{figure}
\centering
\includegraphics[width=8.0cm]{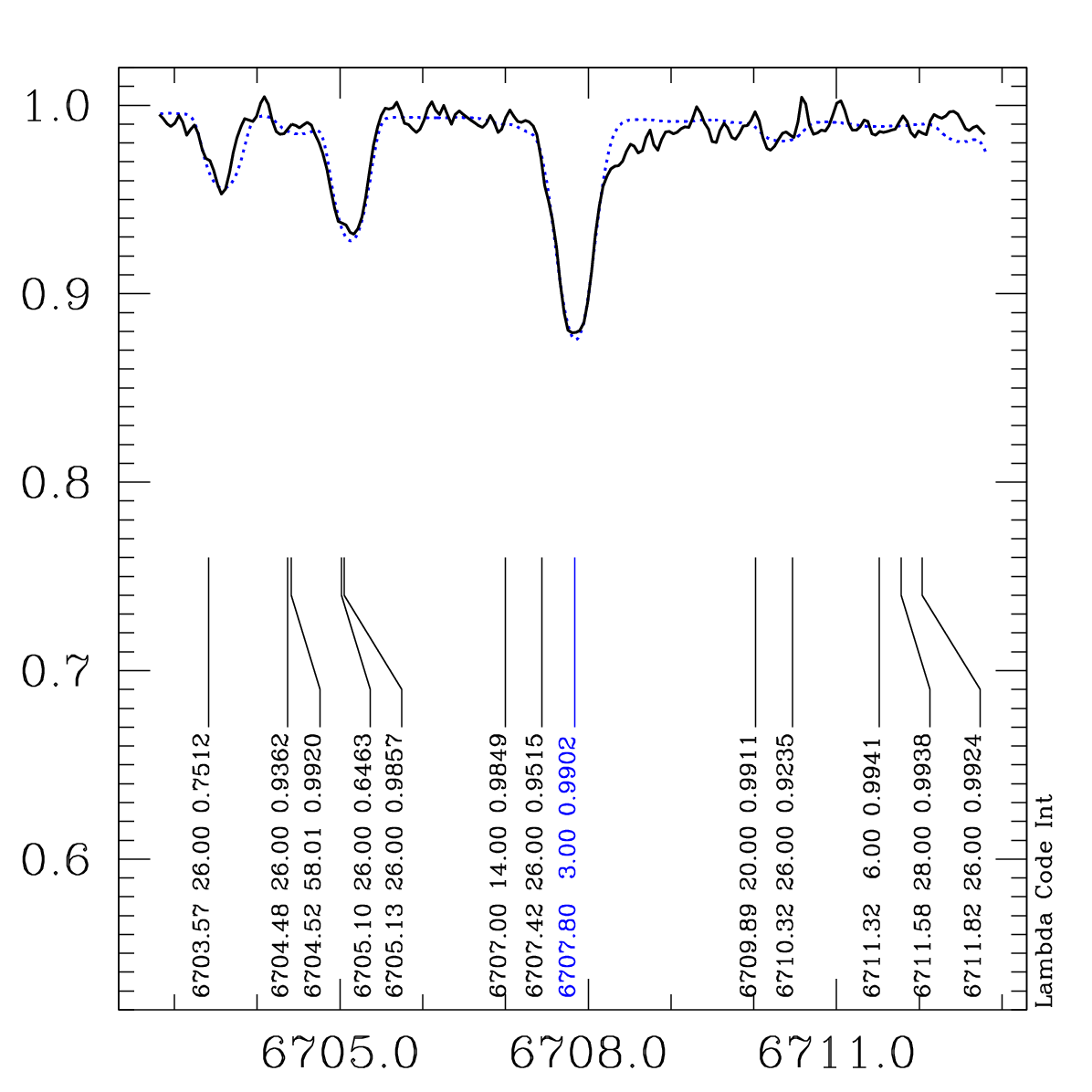}
\caption{Observed spectra (black line) and synthetic spectra (blue dotted line) near the Li line 6707.8 \AA\ in the star CD-33 6615B.
Synthetic lines are indicated showing the wavelength, atomic number, and intensity.}
\label{fig.CD-33.lithium}
\end{figure}

We note that the Li abundance is significantly supersolar in CD-33 6615B ([Li/H]$=$1.72 $\pm$ 0.22 dex).
We present in Fig. \ref{fig.CD-33.lithium} a spectral region near the 
Li line 6707.8 \AA\ in this star.
Observed and synthethic spectra are shown with black and blue dotted lines.
For the synthetic lines, the plot indicates the wavelength, atomic number, and intensity (between 1 and 0).
The lithium line 6707.8 \AA\ was detected in CD-33 6615B but not in their companion HD 87304
(which is a $\lambda$ Boo star, as we will see below).
Interestingly, for the case of the triple system HD 15615,
the Li line was detected in the early-type star HD 15164 \citep[see Fig.5 in ][]{alacoria22}
but not in their $\lambda$ Boo companion HD 15165.
However, we caution that lithium abundance is sensitive to different effects, 
probably not related to the $\lambda$ Boo peculiarity. 
For example, there is a known correlation between lithium content and age. 
Moreover, its abundance depends on other factors such as metallicity in solar-type stars 
\citep[see, for example, Fig. 3 of ][]{martos23}
or even the possible engulfment of a rocky planet \citep{saffe17,soares25}.
We therefore consider that lithium content deserves to be further explored with a larger sample of stars.

\begin{figure*}
\centering
\includegraphics[width=8.0cm]{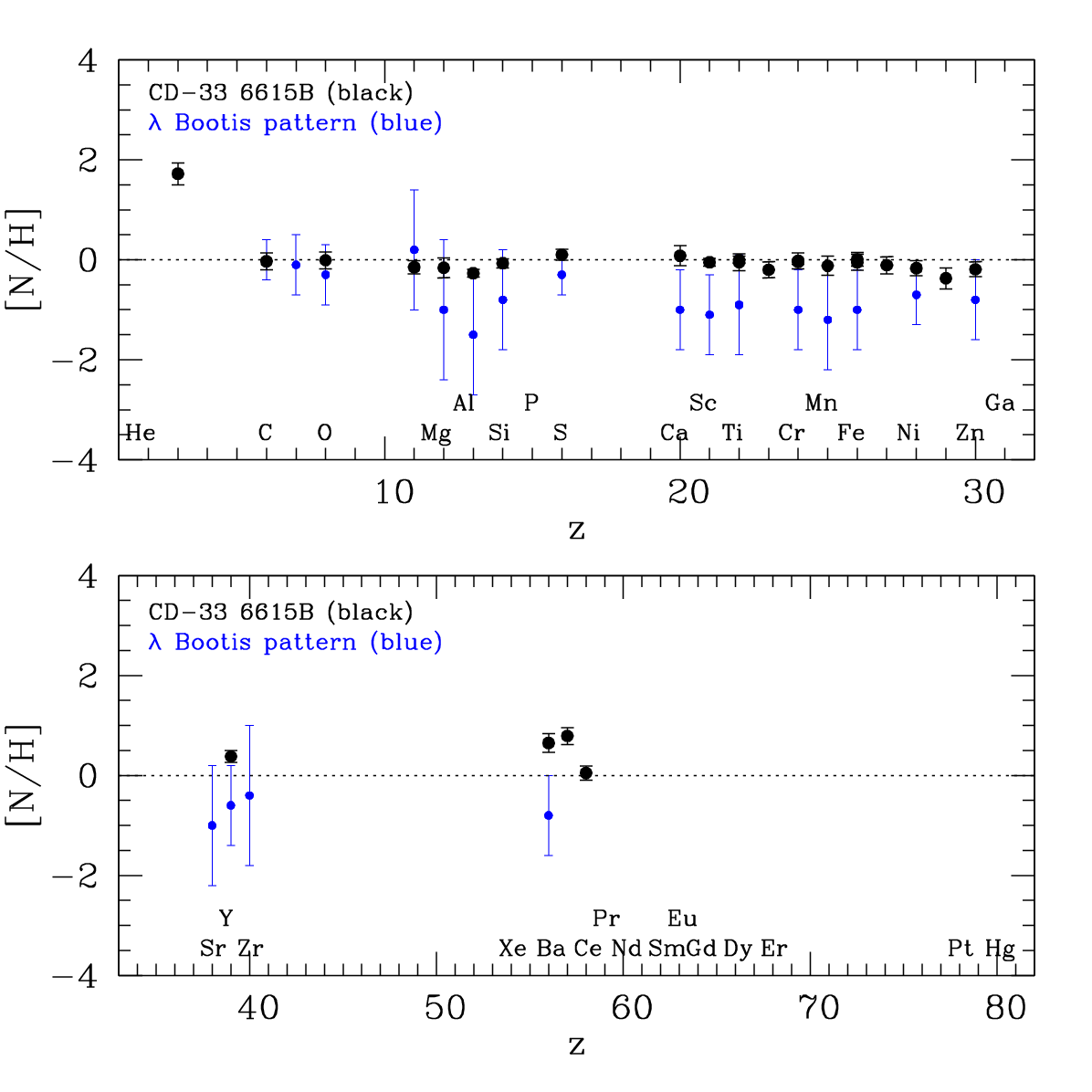}
\includegraphics[width=8.0cm]{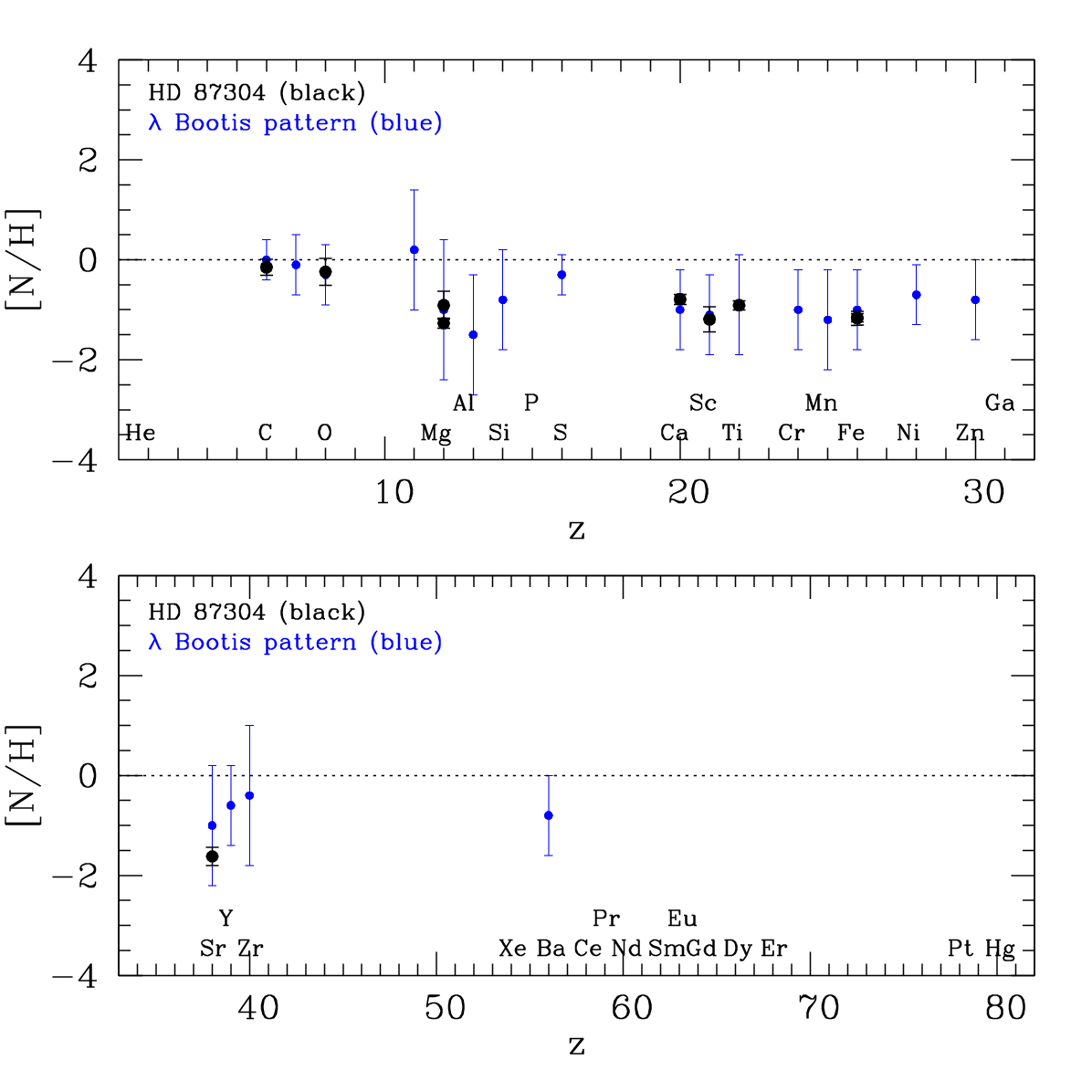}
\caption{Chemical pattern of the stars CD-33 6615B and HD 87304 (left and right panels),
compared to an average pattern of $\lambda$ Boo stars (blue).}
\label{fig.pattern.HD87304}
\end{figure*}

We present in Fig. \ref{fig.pattern.HD87304} the chemical pattern of the stars
CD-33 6615B and HD 87304 (left and right panels), compared to an average pattern of
$\lambda$ Boo stars (blue). For each star, we present two panels, 
corresponding to elements with atomic number z$<$32 and z$>$32.
The error bars of the $\lambda$ Boo pattern show the standard deviation derived from different stars,
while the error bars for our stars correspond to the total error, e$_{tot}$.
From Fig. \ref{fig.pattern.HD87304}, the different chemical composition of both stars is striking.
On one side, CD-33 6615B presents mostly a solar chemical pattern
(for instance, [Fe/H]$=$-0.05 $\pm$ 0.15 dex), with some elements
above solar values (Y, Ba and La).
In particular, the light elements C, O and S present solar abundances.

On the other hand, HD 87304 presents a chemical pattern that agrees with 
those of $\lambda$ Boo stars (see Fig. \ref{fig.pattern.HD87304}).
For instance, C and O present solar or slightly subsolar values 
([C/H]$=$-0.15 $\pm$ 0.16 dex, [O/H]$=$-0.24 $\pm$ 0.27 dex),
which is similar to other $\lambda$ Boo stars. 
Other metals such as Ca, Ti and Fe, present abundances $\sim$1 dex below solar values.
The abundance values confirm the bona fide $\lambda$ Boo nature of this object.
Interestingly, although C and O are almost solar in HD 87304, they seem to be slightly
lower than in CD-33 6615B ([C/H]$=$-0.03 $\pm$ 0.17 dex, [O/H]$=$-0.01 $\pm$ 0.17 dex).
This would suggest that, perhaps, the $\lambda$ Boo phenomena also slightly modifies 
light element abundances (in addition to a stronger effect on heavier species);
however, we caution that NLTE effects could play a role in light elements.
Other metals present a significant difference between both stars
(for example, Fe differs by $\Delta$[Fe/H] $\sim$ 1.12 $\pm$ 0.21 dex).
This is one of the highest differences in metallicity found
between two stars in a binary system \citep[see, for example, ][]{saffe17,saffe24}.

In summary, this binary system is composed by a late-type star with mostly a solar-like composition, 
and an early-type object with a $\lambda$ Boo chemical pattern.
This pair shows that $\lambda$ Boo stars are born with a very different composition,
approximately solar, and reinforces the idea that $\lambda$ Boo stars are Population I objects.
A similar result was obtained by studying the triple system HD 15165 \citep{alacoria22},
the only system reported (up to now) which includes a late-type star with solar-like composition
(HD 15165C) and a $\lambda$ Boo companion (HD 15165).
This is the most likely result of the analysis, however,
we caution that other explanations are also possible (we will come back to this point in section 4.4).

\subsection{Binary system HD 98069 + UCAC4 431-054639}

The star HD 98069 was classified as A9 V kA2mA2 ($\lambda$ Boo) by \citet{murphy20}.
Interestingly, the K2 (Kepler-2) light curve revealed its $\delta$ Scuti nature,
with eigth pulsation peaks exceeding 1 mmag and seven peaks exceeding 0.05 mmag \citep{murphy20,murphy20b}.
This candidate $\lambda$ Boo star (HD 98069) is accompanied by UCAC4 431-054639,
a late-type star separated by 55.24 arcsec or 16148.71 au \citep{el-badry21}.
This binary presents a chance alignment probability of R$=$1.24 10$^{-4}$,
being considered as a high bound probability pair \citep{el-badry21}.
The separation allows us to analyze both stars independently without contamination from its companion.
To our knowledge, there is no detailed abundance determination for HD 98069.

\begin{figure*}
\centering
\includegraphics[width=8.0cm]{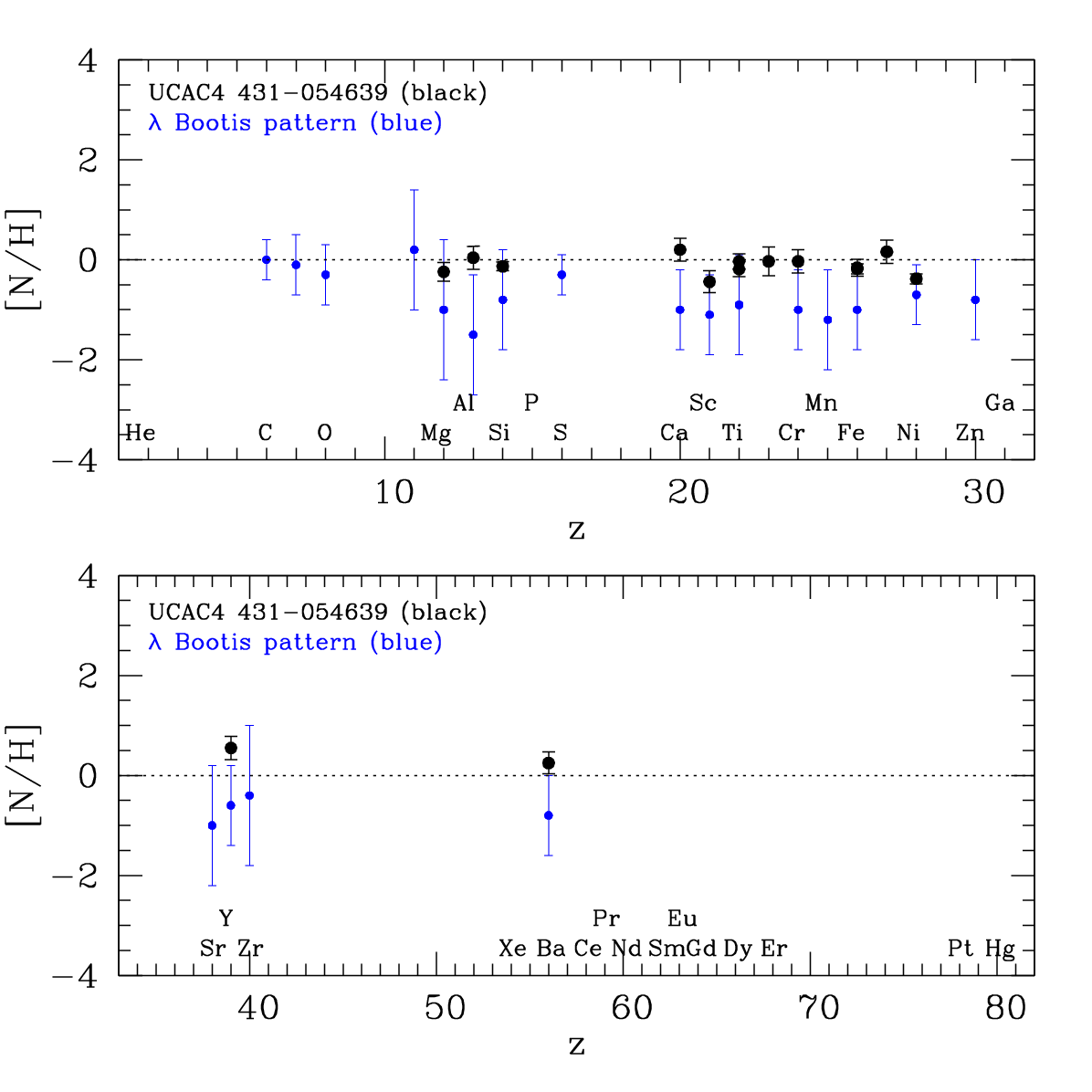}
\includegraphics[width=8.0cm]{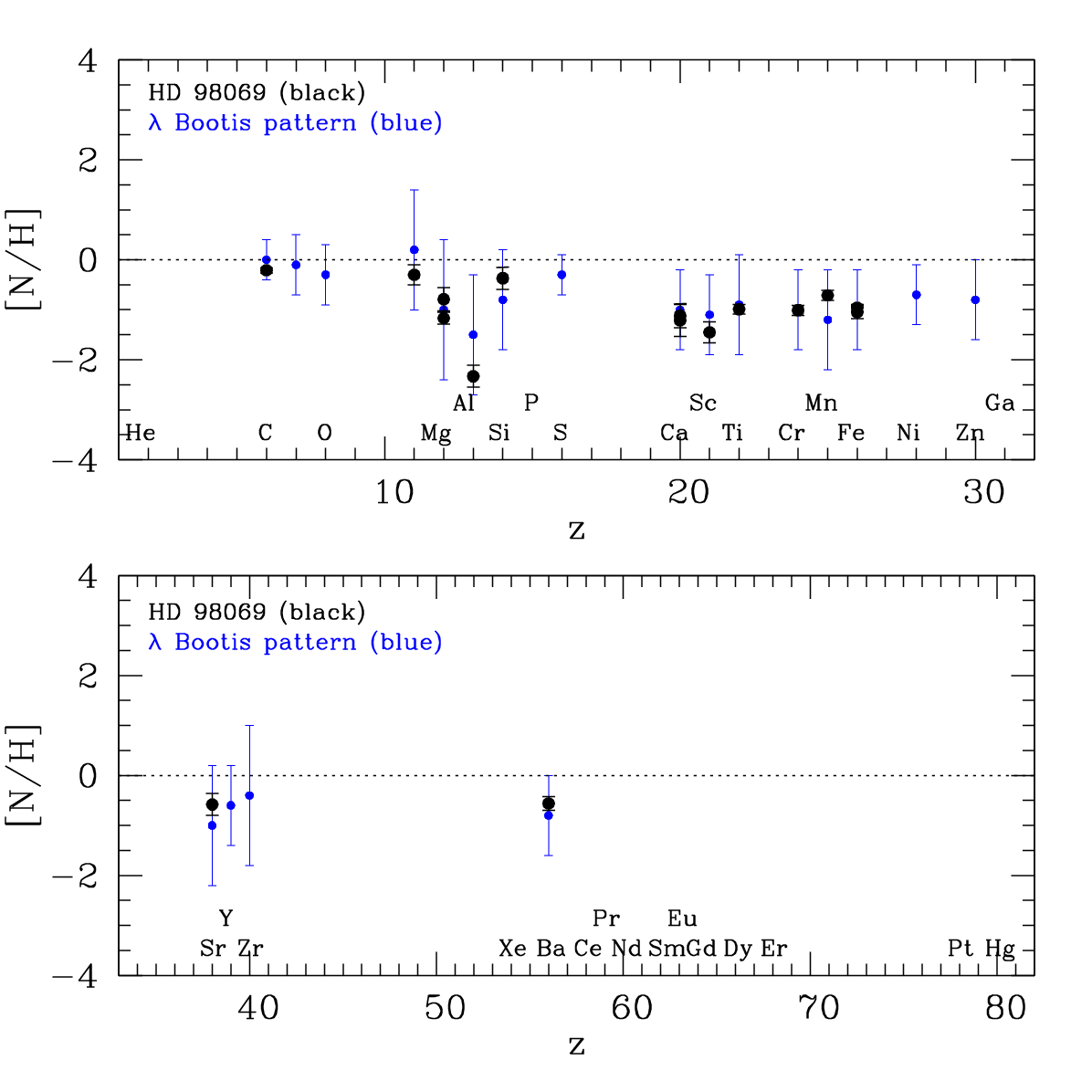}
\caption{Chemical pattern of the stars UCAC4 431-054639 and HD 98069 (left and right panels),
compared to an average pattern of $\lambda$ Boo stars (blue).}
\label{fig.pattern.HD98069}
\end{figure*}

We present in Fig. \ref{fig.pattern.HD98069} the chemical pattern of the stars
UCAC4 431-054639 and HD 98069 (left and right panels), compared to an average pattern of
$\lambda$ Boo stars. The symbols of Fig. \ref{fig.pattern.HD98069} are similar to
those used in the Fig. \ref{fig.pattern.HD87304}.
It is clear from Fig. \ref{fig.pattern.HD98069} that both stars present a significantly
different chemical pattern, similar to the previous binary system.
On one side, UCAC4 431-054639 presents mostly a solar chemical pattern within $\pm\sim$0.20 dex,
with some species showing slightly subsolar (Mg, Sc, Fe) or slightly supersolar (Ca, Y) abundance values. 
For example, for iron we obtained [Fe/H]$=$-0.16 $\pm$ 0.17 dex,
while light-element abundances (CNOS) were not derived.

On the other hand, the chemical pattern of HD 98069 is very different than its stellar companion
(see Fig. \ref{fig.pattern.HD98069}, right panel).
C displays an abundance close to solar, or slightly subsolar ([C/H]$=$ -0.21 $\pm$ 0.05).
Most species are strongly depleted compared to the solar values by $\sim$1 dex or more
(Al, Ca, Ti, Cr, Fe), except perhaps Si which is less depleted ([Si/H]$=$ -0.37 $\pm$ 0.22 dex).
However, we caution that the Si abundance was derived using only one line.
Then, the general pattern of HD 98069 seems to agree with those of $\lambda$ Boo stars.
The metallicity of both stars in this binary system differs by $\Delta$[Fe/H] $\sim$ 0.89 $\pm$ 0.21 dex,
which is also significant although a less extreme difference than the previous binary system.

In summary, this binary system is composed by a late-type star with mostly a solar-like composition, 
and an early-type object with a $\lambda$ Boo-like chemical pattern.

\subsection{Binary system HD 153747 + TYC 7869-2003-1}

The star HD 153747 was classified by \citet{paunzen01b} as hA7mA0 V $\lambda$ Boo,
and later as A7 V kA0mA0 $\lambda$ Boo by \citet{murphy15},
who discussed the membership of this object to the $\lambda$ Boo class.
This object is also reported as $\lambda$ Boo in other literature works \citep{gray17,murphy20}.
\citet{desi14} reported multiperiodicity (periods between 0.96 and 1.2 h) in the light curve of HD 153747,
while \citet{paunzen01b} noted their $\delta$ Sct variability.
For this star, \citet{paunzen02b} reported a depleted metal content of [Z/H]$=$-0.86 $\pm$ 0.20 dex,
estimated using $\Delta$m$_{2}$ from the Geneva 7-color as well as $\Delta$m$_{1}$ from the uvby$\beta$
photometric system.
This candidate $\lambda$ Boo star (HD 153747) is accompanied by TYC 7869-2003-1,
a late-type star separated by 322.64 arcsec or 57131.24 au \citep{el-badry21}.
This pair presents a chance alignment probability of R$=$0.0885,
somewhat higher than previous binary systems, although it is still
considered as a high bound probability pair \citep{el-badry21}.
The separation allows us to analyze both stars independently without contamination from its companion.
To our knowledge, there is no detailed (spectroscopic) abundance determination for any component of this binary system.

\begin{figure*}
\centering
\includegraphics[width=8.0cm]{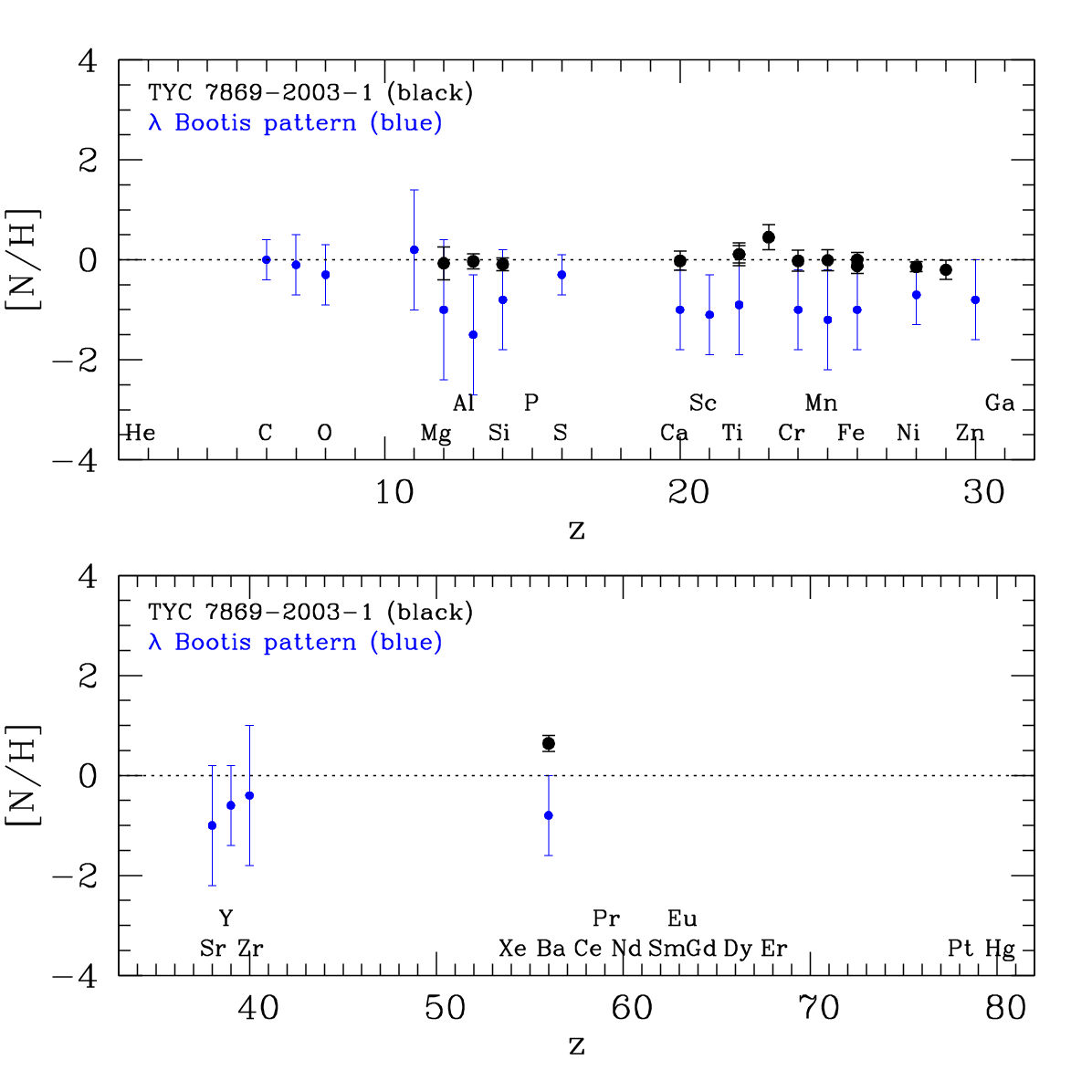}
\includegraphics[width=8.0cm]{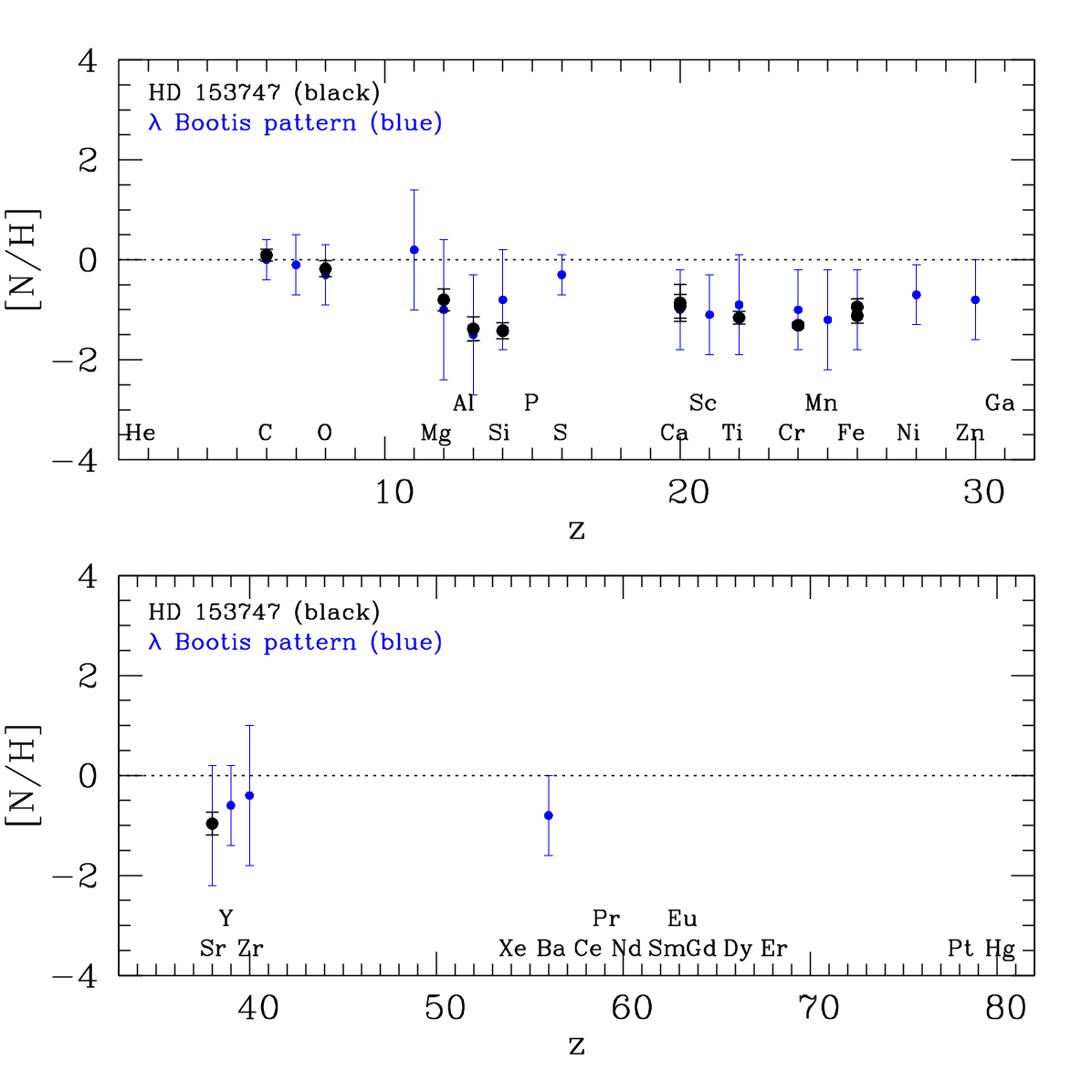}
\caption{Chemical pattern of the stars TYC 7869-2003-1 and HD 153747 (left and right panels),
compared to an average pattern of $\lambda$ Boo stars (blue).}
\label{fig.pattern.HD153747}
\end{figure*}

We present in Fig. \ref{fig.pattern.HD153747} the chemical pattern of the stars
TYC 7869-2003-1 and HD 153747 (left and right panels), compared to an average pattern of
$\lambda$ Boo stars. The symbols of Fig. \ref{fig.pattern.HD153747} are similar to
those used in Figs. \ref{fig.pattern.HD87304} and \ref{fig.pattern.HD98069}.
Both stars present a significantly different chemical pattern, similar to previous binary systems.
TYC 7869-2003-1 displays mostly a solar chemical pattern,
while HD 153747 presents a chemical pattern which closely follows the average pattern 
of $\lambda$ Boo stars. 
HD 153747 displays nearly solar C and O abundances ([C/H]$=$0.09 $\pm$ 0.12 dex, [O/H]$=$-0.18 $\pm$ 0.16 dex),
while other metals show strong depletions of $\sim$1 dex (for example, [Fe/H]$=$-0.94 $\pm$ 0.16 dex).
The metallicity of both stars in this binary system differs by $\Delta$[Fe/H] $\sim$ 0.94 $\pm$ 0.22 dex.

In summary, this binary system is composed by a solar-composition
late-type star, and a $\lambda$ Boo early-type star.

\subsection{The relevance of $\lambda$ Boo stars with late-type companions}

We discuss in this section the importance of finding $\lambda$ Boo stars accompanied by late-type stars.
For the first time, we studied binary systems composed by a late-type object and a candidate $\lambda$ Boo star.
Different authors note that the identification of $\lambda$ Boo stars 
(starting with candidates suggested by spectral classification)
should be followed up with a high-resolution abundance analysis to confirm
their $\lambda$ Boo nature \citep[see e.g.][]{andrievsky02,heiter02,murphy15,gray17,alacoria22}.
In this work, we applied a similar procedure and confirmed the $\lambda$ Boo membership of
three early-type stars (HD 87304, HD 98069 and HD 153747),
as we can see in the Figs. \ref{fig.pattern.HD87304}, \ref{fig.pattern.HD98069} and \ref{fig.pattern.HD153747}.
These three stars could be then considered as bona fide $\lambda$ Boo stars.

We also showed that the late-type companions that belong to these binary systems
present mostly a solar-like composition.
This composition could be used as a proxy for the initial composition of the material
from which the $\lambda$ Boo star formed (under the hypothesis that they are born from the same molecular cloud).
To our knowledge, the starting solar-like composition was only showed in the triple system HD 15165 \citep{alacoria22},
giving us the opportunity to strengthen this result in the present work.
This is an important constraint for any model attempting to explain the $\lambda$ Boo phenomena.
Most scenarios trying to explain the $\lambda$ Boo stars ussually assume a starting solar-like
composition \citep[e.g. ][]{mg09,jura15}.
The present work provides three numerical examples of possible "starting" and "ending" compositions,
which could be further explored to test and constraint numerically formation models of $\lambda$ Boo stars.

In addition, the solar-like composition of the late-type stars is in agreement with the idea that $\lambda$ Boo stars
are Population I objects.
For example, \citet{paunzen14} concluded that it is possible to distinguish $\lambda$ Boo stars from intermediate Population II stars
on the basis of elemental abundances, though not in terms of kinematics.
The authors suggest to use binary systems to strengthen the conclusion that $\lambda$ Boo stars are a distinct population
from the Population II group.
The results of the present work support the idea that $\lambda$ Boo stars belong to Population I,
in agreement with \citet{paunzen14}.
This is the most likely result of the analysis, however,
we caution that other explanations are also possible.
For example, we mentioned that 15\% of A-type stars have companions with periods of 100-1500 days \citep{murphy18}
which would be undetected in the \citet{el-badry21} study. Furthermore, 21\% of those companions are white dwarfs (WDs).
Perhaps, the $\lambda$ Boo star once had a tight orbit with a higher-mass companion that became a red giant, 
transferred mass, and is now an undetected WD, such as those suggested by \citet{murphy18}.
For example, \citet{van-winckel95} presented five extremely Fe-deficient post-AGB binary systems with
orbital periods from one to a few years, suggesting that mass transfer has occurred in these systems.
\citet{paunzen98} suggested that a similar mass-transfer might also be an additional mechanism to be considered
for the $\lambda$ Boo phenomenon.
Then, in our binary systems, the cool star could have been captured by the hypothetical tight binary and be of any age.

In general, the present results are in agreement with those previously obtained in the triple system HD 15165 \citep{alacoria22},
in which a $\lambda$ Boo star (HD 15165) is accompanied by a late-type star with solar-like
composition (HD 15165C).
To our knowledge, $\lambda$ Boo stars were detected in binary (and multiple) systems accompanied
by early-type stars \citep[e.g. ][]{paunzen12a,paunzen12b,alacoria22},
by late-type companions \citep[this work and ][]{alacoria22},
and even one case accompanied by a Brown Dwarf \citep[$\zeta$ Del, detected by ][]{saffe21}.
However, no bona fide $\lambda$ Boo star have been detected in open clusters
\citep[e.g. ][]{paunzen01a,paunzen01b,gray-corbally02},
including searches in different intermediate-age open clusters.
Other chemically peculiar stars (such as Am or Ap) were detected in the same clusters \citep[e.g. ][]{gray-corbally02}.
It would be valuable to detect $\lambda$ Boo stars in open clusters, which could also be used
to study their origin.

The work of \citet{jura15} mentioned the possibility that $\lambda$ Boo stars could
originate by accreting the winds from late-type stellar companions (their Section 4.1).
\citet{jura15} considered that only a small fraction of the material from the late-type stars
could reach the $\lambda$ Boo star and then ruled out this scenario
(instead, the author proposed winds from hot-Jupiter planets as a plausible mechanism).
In the present work, we analyzed binary systems incluing $\lambda$ Boo stars
and late-type companions. However, the three binary systems analyzed here present wide separations
(from 3160 au to 57131 au, see Table \ref{table.LamBoo.binaries}), making unlikely the suggested mechanism.
On the other hand, there are no planets reported orbiting the $\lambda$ Boo stars studied in this work.





\section{Concluding remarks}

In the present work, we cross-matched an homogeneous list of candidate $\lambda$ Boo stars
with a recent catalog of resolved binaries detected with Gaia eDR3, with the aim of finding
$\lambda$ Boo stars as members of multiple systems.
Then, we performed a detailed abundance determination of three of these binary systems.
The main results of this work are as follows:

-We present a group of 19 newly identified binary systems that contains
a candidate $\lambda$ Boo star (see Table \ref{table.LamBoo.binaries}).
The new systems allow to $\sim$duplicate 
the number of $\lambda$ Boo stars currently known in multiple systems.
This important group could be used in further studies of $\lambda$ Boo stars.

-For the first time, we performed a detailed abundance analysis of three binary systems
including a candidate $\lambda$ Boo star and a late-type companion. 
We confirmed the true $\lambda$ Boo nature of the three early-type stars,
and obtained mostly a solar-like composition for the late-type components.
In particular, the binary system HD 87304 + CD-33 6615B presents a mutual metallicity
difference of $\Delta$[Fe/H] $\sim$ 1.12 $\pm$ 0.21 dex, one of the highest differences found in 
a binary system.

-Adopting as a proxy the chemical composition of the late-type stars,
we showed that the three $\lambda$ Boo stars were initially born with a solar-like 
composition. This is an important constraint for any scenario trying to
explain the origin of $\lambda$ Boo stars.
The present work provides three numerical examples of possible "starting" and "ending" compositions
to test the models.

-Finally, the solar-like composition of the late-type stars supports the idea
that $\lambda$ Boo stars are Population I objects.
This is in agreement with the suggestions of previous works \citep{paunzen14,alacoria22}.
However, we caution that other explanations are also possible.

The present work shows the importance of finding $\lambda$ Boo stars that belong
to multiple systems. The stars used in this work correspond mostly to southern $\lambda$ Boo stars.
We encourage to perform a similar work on additional binary systems,
and to expand the list of homogeneous $\lambda$ Boo candidates to the North hemisphere.

\begin{acknowledgements}
We thank the referee for constructive comments that improved the paper.
The authors thank Dr. R. Kurucz for making their codes available to us.
CS acknowledge financial support from CONICET (Argentina) through grant PIP 11220210100048CO
and the National University of San Juan (Argentina) through grant CICITCA 21/E1235.
IRAF is distributed by the National Optical Astronomical Observatories, 
which is operated by the Association of Universities for Research in Astronomy, Inc., under a cooperative agreement
with the National Science Foundation.
Based on data acquired at Complejo Astron\'omico El Leoncito, operated under agreement between
the Consejo Nacional de Investigaciones Cient\'ificas y T\'ecnicas de la Rep\'ublica Argentina and
the National Universities of La Plata, C\'ordoba and San Juan.

\end{acknowledgements}

\begin{appendix}

\section{Chemical abundances}

 In this section, we present the chemical abundances and their corresponding errors.
The total error e$_{tot}$ was derived as the quadratic sum of the line-to-line dispersion e$_{1}$
(estimated as $\sigma/\sqrt{n}$ , where $\sigma$ is the standard deviation)
and the error in the abundances (e$_{2}$, e$_{3}$, and e$_{4}$) when varying T$_{\rm eff}$, $\log g$,
and v$_\mathrm{micro}$ by their corresponding uncertainties\footnote{We adopt a minimum of 0.01 dex for the errors e$_{2}$, e$_{3}$, and e$_{4}$.}.
For chemical species with only one line, we adopted as $\sigma$ the standard deviation of iron lines.
Abundance tables show the average abundance and the total error e$_{tot}$, together with
the errors e$_{1}$ to e$_{4}$.

\begin{table}
\centering
\caption{Chemical abundances for the star HD 87304.}
\begin{tabular}{lrcccc}
\hline
\hline
Species    & [X/H] $\pm$ e$_{tot}$ & e$_{1}$ & e$_{2}$ & e$_{3}$ & e$_{4}$ \\
\hline
C I     & -0.15 $\pm$ 0.16 & 0.16 & 0.01 & 0.01 & 0.02 \\ 
O I     & -0.24 $\pm$ 0.27 & 0.25 & 0.04 & 0.01 & 0.09 \\ 
Mg I    & -1.27 $\pm$ 0.10 & 0.06 & 0.07 & 0.02 & 0.04 \\ 
Mg II   & -0.91 $\pm$ 0.28 & 0.25 & 0.12 & 0.03 & 0.03 \\ 
Ca II   & -0.79 $\pm$ 0.10 & 0.04 & 0.09 & 0.01 & 0.02 \\ 
Sc II   & -1.19 $\pm$ 0.25 & 0.25 & 0.05 & 0.02 & 0.03 \\ 
Ti II   & -0.91 $\pm$ 0.09 & 0.04 & 0.05 & 0.02 & 0.05 \\ 
Fe I    & -1.17 $\pm$ 0.14 & 0.07 & 0.11 & 0.01 & 0.04 \\ 
Fe II   & -1.16 $\pm$ 0.08 & 0.06 & 0.04 & 0.01 & 0.04 \\ 
Sr II   & -1.62 $\pm$ 0.18 & 0.05 & 0.15 & 0.02 & 0.08 \\ 
\hline
\end{tabular}
\label{tab.abunds.HD87304}
\end{table}

\begin{table}
\centering
\caption{Chemical abundances for the star CD-33 6615B.}
\begin{tabular}{lrcccc}
\hline
\hline
Species    & [X/H] $\pm$ e$_{tot}$ & e$_{1}$ & e$_{2}$ & e$_{3}$ & e$_{4}$ \\
\hline
Li I    &  1.72 $\pm$ 0.22 & 0.12 & 0.18 & 0.01 & 0.01 \\ 
C I     & -0.03 $\pm$ 0.17 & 0.02 & 0.16 & 0.03 & 0.01 \\ 
O I     & -0.01 $\pm$ 0.17 & 0.01 & 0.16 & 0.02 & 0.01 \\ 
Na I    & -0.15 $\pm$ 0.13 & 0.12 & 0.04 & 0.01 & 0.01 \\ 
Mg I    & -0.16 $\pm$ 0.20 & 0.10 & 0.18 & 0.01 & 0.01 \\ 
Al I    & -0.27 $\pm$ 0.08 & 0.04 & 0.07 & 0.01 & 0.01 \\ 
Si I    & -0.07 $\pm$ 0.09 & 0.05 & 0.07 & 0.01 & 0.01 \\ 
S I     &  0.10 $\pm$ 0.11 & 0.03 & 0.11 & 0.02 & 0.01 \\ 
Ca I    &  0.08 $\pm$ 0.20 & 0.11 & 0.15 & 0.01 & 0.05 \\ 
Sc II   & -0.05 $\pm$ 0.08 & 0.06 & 0.03 & 0.02 & 0.03 \\ 
Ti I    & -0.05 $\pm$ 0.17 & 0.03 & 0.17 & 0.01 & 0.01 \\  
Ti II   & -0.00 $\pm$ 0.07 & 0.05 & 0.02 & 0.02 & 0.04 \\  %
V I     & -0.20 $\pm$ 0.16 & 0.09 & 0.13 & 0.01 & 0.02 \\ 
Cr I    & -0.02 $\pm$ 0.16 & 0.02 & 0.16 & 0.01 & 0.02 \\  
Cr II   & -0.04 $\pm$ 0.05 & 0.03 & 0.02 & 0.01 & 0.03 \\  
Mn I    & -0.12 $\pm$ 0.19 & 0.12 & 0.14 & 0.01 & 0.05 \\ 
Fe I    & -0.05 $\pm$ 0.16 & 0.01 & 0.15 & 0.01 & 0.04 \\ 
Fe II   &  0.01 $\pm$ 0.14 & 0.03 & 0.12 & 0.01 & 0.07 \\ 
Co I    & -0.11 $\pm$ 0.17 & 0.12 & 0.12 & 0.01 & 0.01 \\ 
Ni I    & -0.17 $\pm$ 0.15 & 0.03 & 0.14 & 0.01 & 0.03 \\ 
Cu I    & -0.37 $\pm$ 0.21 & 0.12 & 0.17 & 0.01 & 0.01 \\ 
Zn I    & -0.19 $\pm$ 0.15 & 0.06 & 0.11 & 0.01 & 0.09 \\ 
Y II    &  0.38 $\pm$ 0.12 & 0.12 & 0.01 & 0.01 & 0.01 \\ 
Ba II   &  0.65 $\pm$ 0.19 & 0.12 & 0.09 & 0.01 & 0.12 \\ 
La II   &  0.79 $\pm$ 0.17 & 0.12 & 0.12 & 0.02 & 0.03 \\ 
Ce II   &  0.05 $\pm$ 0.14 & 0.12 & 0.07 & 0.02 & 0.01 \\ 
\hline
\end{tabular}
\label{tab.abunds.CD-33}
\end{table}

\begin{table}
\centering
\caption{Chemical abundances for the star HD 98069.}
\begin{tabular}{lrcccc}
\hline
\hline
Species    & [X/H] $\pm$ e$_{tot}$ & e$_{1}$ & e$_{2}$ & e$_{3}$ & e$_{4}$ \\
\hline
C I     & -0.21 $\pm$ 0.05 & 0.04 & 0.04 & 0.01 & 0.01 \\ 
Na I    & -0.30 $\pm$ 0.20 & 0.18 & 0.08 & 0.02 & 0.01 \\ 
Mg I    & -1.17 $\pm$ 0.12 & 0.06 & 0.08 & 0.02 & 0.07 \\ 
Mg II   & -0.79 $\pm$ 0.23 & 0.18 & 0.14 & 0.02 & 0.03 \\ 
Al I    & -2.33 $\pm$ 0.22 & 0.18 & 0.13 & 0.01 & 0.01 \\ 
Si II   & -0.37 $\pm$ 0.22 & 0.18 & 0.13 & 0.02 & 0.01 \\ 
Ca I    & -1.21 $\pm$ 0.32 & 0.18 & 0.21 & 0.03 & 0.16 \\ 
Ca II   & -1.12 $\pm$ 0.24 & 0.18 & 0.13 & 0.07 & 0.04 \\ 
Sc II   & -1.45 $\pm$ 0.21 & 0.18 & 0.10 & 0.01 & 0.04 \\ 
Ti II   & -0.99 $\pm$ 0.10 & 0.06 & 0.04 & 0.01 & 0.06 \\ 
Cr II   & -1.01 $\pm$ 0.10 & 0.10 & 0.02 & 0.01 & 0.01 \\ 
Mn I    & -0.71 $\pm$ 0.10 & 0.08 & 0.06 & 0.01 & 0.01 \\ 
Fe I    & -1.05 $\pm$ 0.13 & 0.04 & 0.11 & 0.02 & 0.03 \\ 
Fe II   & -0.96 $\pm$ 0.07 & 0.05 & 0.04 & 0.01 & 0.03 \\ 
Sr II   & -0.58 $\pm$ 0.22 & 0.04 & 0.13 & 0.01 & 0.18 \\ 
Ba II   & -0.56 $\pm$ 0.14 & 0.02 & 0.13 & 0.01 & 0.05 \\ 
\hline
\end{tabular}
\label{tab.abunds.HD98069}
\end{table}

\begin{table}
\centering
\caption{Chemical abundances for the star UCAC4 431-054639.}
\begin{tabular}{lrcccc}
\hline
\hline
Species    & [X/H] $\pm$ e$_{tot}$ & e$_{1}$ & e$_{2}$ & e$_{3}$ & e$_{4}$ \\
\hline
Mg I    & -0.24 $\pm$ 0.19 & 0.09 & 0.17 & 0.01 & 0.01 \\ 
Al I    &  0.04 $\pm$ 0.23 & 0.20 & 0.13 & 0.01 & 0.01 \\ 
Si I    & -0.13 $\pm$ 0.09 & 0.07 & 0.05 & 0.01 & 0.01 \\ 
Ca I    &  0.20 $\pm$ 0.23 & 0.12 & 0.20 & 0.01 & 0.03 \\ 
Sc II   & -0.44 $\pm$ 0.22 & 0.20 & 0.08 & 0.03 & 0.06 \\ 
Ti I    & -0.19 $\pm$ 0.15 & 0.03 & 0.14 & 0.01 & 0.04 \\  
Ti II   & -0.03 $\pm$ 0.15 & 0.06 & 0.12 & 0.05 & 0.03 \\   
V I     & -0.03 $\pm$ 0.29 & 0.20 & 0.21 & 0.01 & 0.01 \\ 
Cr I    & -0.03 $\pm$ 0.23 & 0.11 & 0.20 & 0.03 & 0.05 \\ 
Fe I    & -0.16 $\pm$ 0.17 & 0.04 & 0.16 & 0.01 & 0.02 \\ 
Fe II   & -0.18 $\pm$ 0.10 & 0.05 & 0.07 & 0.05 & 0.01 \\ 
Co I    &  0.16 $\pm$ 0.23 & 0.20 & 0.12 & 0.01 & 0.01 \\ 
Ni I    & -0.38 $\pm$ 0.10 & 0.06 & 0.08 & 0.01 & 0.01 \\ 
Y II    &  0.55 $\pm$ 0.23 & 0.20 & 0.11 & 0.03 & 0.06 \\ 
Ba II   &  0.25 $\pm$ 0.22 & 0.20 & 0.02 & 0.03 & 0.09 \\ 
\hline
\end{tabular}
\label{tab.abunds.UCAC4}
\end{table}

\begin{table}
\centering
\caption{Chemical abundances for the star HD 153747.}
\begin{tabular}{lrcccc}
\hline
\hline
Species    & [X/H] $\pm$ e$_{tot}$ & e$_{1}$ & e$_{2}$ & e$_{3}$ & e$_{4}$ \\
\hline
C I     &  0.09 $\pm$ 0.12 & 0.02 & 0.12 & 0.01 & 0.01 \\ 
O I     & -0.18 $\pm$ 0.16 & 0.16 & 0.02 & 0.01 & 0.01 \\ 
Mg I    & -0.80 $\pm$ 0.22 & 0.11 & 0.19 & 0.01 & 0.04 \\ 
Al I    & -1.38 $\pm$ 0.24 & 0.16 & 0.17 & 0.01 & 0.01 \\ 
Si II   & -1.42 $\pm$ 0.16 & 0.16 & 0.01 & 0.01 & 0.01 \\ 
Ca I    & -0.86 $\pm$ 0.37 & 0.16 & 0.33 & 0.02 & 0.04 \\ 
Ca II   & -0.93 $\pm$ 0.24 & 0.16 & 0.18 & 0.01 & 0.01 \\ 
Ti II   & -1.16 $\pm$ 0.13 & 0.02 & 0.13 & 0.01 & 0.01 \\ 
Cr II   & -1.31 $\pm$ 0.05 & 0.04 & 0.02 & 0.01 & 0.01 \\ 
Fe I    & -0.94 $\pm$ 0.16 & 0.04 & 0.13 & 0.01 & 0.08 \\ 
Fe II   & -1.12 $\pm$ 0.15 & 0.09 & 0.12 & 0.01 & 0.02 \\ 
Sr II   & -0.96 $\pm$ 0.23 & 0.03 & 0.23 & 0.01 & 0.04 \\ 
\hline
\end{tabular}
\label{tab.abunds.HD153747}
\end{table}

\begin{table}
\centering
\caption{Chemical abundances for the star TYC 7869-2003-1.}
\begin{tabular}{lrcccc}
\hline
\hline
Species    & [X/H] $\pm$ e$_{tot}$ & e$_{1}$ & e$_{2}$ & e$_{3}$ & e$_{4}$ \\
\hline
Mg I    & -0.07 $\pm$ 0.33 & 0.15 & 0.29 & 0.01 & 0.01 \\ 
Al I    & -0.03 $\pm$ 0.15 & 0.15 & 0.04 & 0.01 & 0.01 \\ 
Si I    & -0.09 $\pm$ 0.13 & 0.13 & 0.01 & 0.01 & 0.01 \\ 
Ca I    & -0.02 $\pm$ 0.19 & 0.16 & 0.11 & 0.01 & 0.01 \\ 
Ti I    &  0.11 $\pm$ 0.17 & 0.08 & 0.15 & 0.01 & 0.02 \\ 
Ti II   &  0.11 $\pm$ 0.23 & 0.22 & 0.05 & 0.01 & 0.01 \\ 
V I     &  0.45 $\pm$ 0.25 & 0.08 & 0.24 & 0.01 & 0.01 \\ 
Cr I    & -0.02 $\pm$ 0.21 & 0.15 & 0.15 & 0.01 & 0.02 \\ 
Mn I    & -0.01 $\pm$ 0.21 & 0.13 & 0.16 & 0.01 & 0.01 \\ 
Fe I    &  0.00 $\pm$ 0.15 & 0.03 & 0.14 & 0.01 & 0.01 \\ 
Fe II   & -0.12 $\pm$ 0.15 & 0.03 & 0.15 & 0.01 & 0.01 \\ 
Ni I    & -0.14 $\pm$ 0.10 & 0.06 & 0.07 & 0.01 & 0.01 \\ 
Cu I    & -0.20 $\pm$ 0.19 & 0.15 & 0.11 & 0.01 & 0.03 \\ 
Ba II   &  0.64 $\pm$ 0.16 & 0.15 & 0.06 & 0.01 & 0.02 \\ 
\hline
\end{tabular}
\label{tab.abunds.TYC7869}
\end{table}

\end{appendix}

\end{document}